\documentclass{article}

\usepackage{arxiv}

\usepackage{amssymb}
\usepackage{amsmath}
\usepackage{lineno}
\usepackage{textcomp}
\usepackage{float}  
\usepackage{caption}
\usepackage{subcaption}
\usepackage{indentfirst}
\usepackage{enumerate}
\usepackage{microtype}
\usepackage{chemformula}
\usepackage{setspace}
\usepackage{yfonts,color}
\usepackage{siunitx}
\usepackage{comment}
\usepackage{multirow}
\usepackage{varioref}
\usepackage[bottom]{footmisc}
\usepackage{wrapfig}
\usepackage{float}
\usepackage{type1cm}
\usepackage{chngcntr}
\usepackage[nottoc, notlof, notlot]{tocbibind}
\usepackage{hyperref}
\usepackage{subfiles}
\usepackage{soul} 
\usepackage{graphicx}

\begin{document}


\title{Kaonic Copper and Fluorine Absolute Yields Measurement with a CZT-based Detection System at DA$\Phi$NE}

\maketitle

Francesco Artibani$^{1,2,*}$,
Simone Manti$^1$,
Leonardo Abbene$^{3,1}$,
Antonino Buttacavoli$^{3,1}$,
Manuele Bettelli$^4$,
Gaetano Gerardi$^{3}$,
Fabio Principato$^{3,1}$,
Andrea Zappettini$^4$,
Massimiliano Bazzi$^1$,
Giacomo Borghi$^{5,6}$,
Damir Bosnar$^{7,1}$,
Mario Bragadireanu$^8$,
Marco Carminati$^{5,6}$,
Alberto Clozza$^1$,
Francesco Clozza$^{1,10}$,
Raffaele Del Grande$^{11,1}$,
Luca De Paolis$^1$,
Carlo Fiorini$^{5,6}$,
Ivica Friscic$^7$,
Carlo Guaraldo$^1$\textsuperscript{\textdagger},
Mihail Iliescu$^1$,
Masahiko Iwasaki$^{14}$,
Aleksander Khreptak$^{12,13,1}$,
Johann Marton$^9$,
Pawel Moskal$^{12,13}$,
Fabrizio Napolitano$^{15,16,1}$,
Hiroaki Ohnishi$^{17}$,
Kristian Piscicchia$^{18,1}$,
Francesco Sgaramella$^1$,
Michal Silarski$^{12}$,
Diana Laura Sirghi$^{1,8,18}$,
Florin Sirghi$^{1,8}$,
Magdalena Skurzok$^{12,13,1}$,
Antonio Spallone$^1$,
Kairo Toho$^{17,1}$,
Oton Vazquez Doce$^1$,
Johann Zmeskal\textsuperscript{\textdagger},
Catalina Curceanu$^1$,
Alessandro Scordo$^1$

{$^1$Laboratori Nazionali di Frascati, INFN, Via E. Fermi 54, 00044 Frascati, Italy} \\
\\{$^2$Dipartimento di Matematica e Fisica, Università di Roma Tre, Via della Vasca Navale 84, 00146 Roma, Italy} \\
\\{$^3$Dipartimento di Fisica e Chimica - Emilio Segrè, Università di Palermo, Viale Delle Scienze Edificio 18, 90128 Palermo, Italy} \\
\\{$^4$Istituto Materiali per l’Elettronica e il Magnetismo, Consiglio Nazionale delle Ricerche, Parco Area delle Scienze 37/A, 43124 Parma, Italy}\\
\\{$^5$Dipartimento di Elettronica, Informazione e Bioingegneria, Politecnico di Milano, Piazza Leonardo da Vinci 32, 20133 Milano, Italy}\\
\\{$^6$INFN Sezione di Milano, Via Giovanni Celoria 16, 20133 Milano, Italy}\\
\\{$^7$Department of Physics, Faculty of Science, University of Zagreb, Bijenička cesta 32, 10000 Zagreb, Croatia}\\
\\{$^8$Horia Hulubei National Institute of Physics and Nuclear Engineering (IFIN-HH), No. 30, Reactorului Street, 077125 Magurele, Ilfov, Romania}\\
\\{$^9$Atominstitut, Technical University Vienna, Stadionallee 2, 1020 Vienna, Austria}\\
\\{$^{10}$Dipartimento di Scienze Matematiche Fisiche e Naturali, Università degli Studi di Roma Tor Vergata, Via Cracovia 90, 00133 Roma, Italy}\\
\\{$^{11}$Faculty of Nuclear Sciences and Physical Engineering, Czech Technical University in Prague, Břehová 7, 115 19 Prague, Czech Republic}\\
\\{$^{12}$Faculty of Physics, Astronomy, and Applied Computer Science, Jagiellonian University, Lojasiewicza 11, 30-348 Krakow, Poland}\\
\\{$^{13}$Centre for Theranostics, Jagiellonian University, Kopernika 40, 31-501 Krakow, Poland}\\
\\{$^{14}$RIKEN, 2-1 Hirosawa, Wako, Saitama 351-0198, Japan}\\
\\{$^{15}$Via A. Pascoli 06123, Perugia (PG), Italy, Dipartimento di Fisica e Geologia, Università degli studi di Perugia}\\
\\{$^{16}$INFN Sezione di Perugia, Via A. Pascoli – 06123 Perugia – Italia}\\
\\{$^{17}$Research Center for Accelerator and Radioisotope Science (RARIS), Tohoku University, 1-2-1 Mikamine, Taihaku-ku, 982-0826 Sendai, Japan}\\
\\{$^{18}$Centro Ricerche Enrico Fermi - Museo Storico della Fisica e Centro Studi e Ricerche “Enrico Fermi”, Via Panisperna 89A, 00184 Roma, Italy}\\
\\{\textsuperscript{\textdagger} Deceased.}\\

{$^*$Author to whom any correspondence should be addressed.}

{francesco.artibani@infn.it, francesco.artibani@uniroma3.it}


\bigskip

\begin{abstract}

\noindent In this work, new measurements of absolute X-ray yields for several transitions in kaonic copper and, for the first time, in kaonic fluorine are reported. The data were collected by the SIDDHARTA-2 collaboration at the DA$\Phi$NE collider using a novel room-temperature Cadmium Zinc Telluride (CZT) detection system. Detection efficiencies were evaluated through a dedicated Geant4 Monte Carlo simulation of the full experimental setup, enabling the extraction of absolute yields per stopped kaon.

\noindent The measured yields exhibit a systematic dependence on the principal quantum number, reflecting the interplay between radiative transitions, Auger de-excitation, and strong-interaction-induced nuclear capture. In kaonic fluorine, a suppression of the 4$\to$3 transition yield relative to higher-n transitions is observed, providing evidence for the onset of strong-interaction effects already at the $n=4$ level. From this behaviour, a conservative lower limit on the corresponding strong-interaction width is derived.

\noindent These results provide new quantitative constraints for cascade models of exotic atoms and extend experimental access to intermediate atomic levels where strong-interaction effects are not directly observable via level shifts and widths. They also establish CZT-based detection as a powerful and versatile approach for high-resolution X-ray spectroscopy of kaonic atoms in collider environments.

\end{abstract}


\section{Introduction}

\noindent Kaonic atoms provide a relevant experimental framework to investigate strong interactions involving strangeness at low energies \cite{Friedman:1994hx, Batty:1997zp, Cieply:2016jby, Curceanu:2019uph, Artibani:2024kop}. These exotic systems are formed when a negatively charged kaon stops inside a material and binds to a nucleus via the electromagnetic interaction. The capture occurs in a highly excited state, when the wavefunction overlap of the kaon with the outermost electron orbit of the target atom becomes significant \cite{Zmeskal:2008zz}.

\noindent The kaon subsequently undergoes an atomic cascade de-excitation, governed by the competition between three competing processes: Auger electron emission, which dominates at high principal quantum numbers where the kaon wavefunction is far from the nucleus; radiative X-ray emission, which becomes increasingly important at lower quantum numbers as the kaon approaches the nucleus; strong interaction induced nuclear capture, which becomes competitive only in the very last atomic levels when the kaon is sufficiently close to the nucleons to enable strong interaction between the meson and the hadrons. The relative importance of these three processes encodes information about both atomic and molecular dynamics \cite{Aramaki:2013fqa} and the low-energy kaon-nucleon (K-N) and kaon-multinucleon (K-multiN) interaction \cite{Ericson:1970ij}.

\noindent Kaonic atom spectroscopy offers three complementary observables to constrain theoretical models of the strong interaction with strangeness \cite{Friedman:1994hx,Curceanu:2019uph, Curceanu:2026zjg}. The first two observables consist of the energy shifts ($\epsilon$) and widths ($\Gamma$) of the last observable X-ray transitions, which provide direct information on the strong interaction in the final atomic levels. The third observable is the measurement of radiative yields of two consecutive transition, which, in some case, can provide access to the intrinsic widths $\Gamma_u$ of levels where the strong interaction is too weak to be directly resolved by conventional X-ray detectors \cite{Friedman:2013dva}. Both quantities serve as unique observables for testing theoretical models of K-N and K-multiN interactions in the nuclear medium \cite{Ericson:1970ij, Obertova:2025rso}.

\noindent Moreover, measuring X-ray yields for transitions where the strong interaction is negligible, is essential for benchmarking cascade models. These models \cite{ Aramaki:2013fqa, Markushin:2001tv, Koike:2007uye} aim to predict the de-excitation path of kaon through the atomic levels. To do so, they must account for complex and poorly understood processes, such as electron refilling, collisional effects, and the initial population distribution of atomic states. As a result, cascade models still suffer from significant uncertainties and rely on parameters previously constrained by direct yield measurements in exotic atoms \cite{Aramaki:2013fqa}.

\noindent Over the last decades, a new generation of kaonic atoms experiments performed at facilities such as DA$\Phi$NE, KEK, and J-PARC has significantly improved the spectroscopic information in the light kaonic atom mass region \cite{Ishiwatari:2004cf, SIDDHARTA:2013ftw, Bazzi:2016rlt, Sgaramella:2024mnb, Sgaramella:2024klx}, enabling high-precision measurements of X-ray transitions. The intermediate and high mass region, however, still remains poorly explored, particularly regarding absolute yield measurements.

\noindent In this paper, the first measurements ever of absolute X-ray yields for kaonic fluorine (Z=9) transitions in teflon, and new measurements of kaonic copper transitions (Z=29), performed by the SIDDHARTA-2 collaboration, are presented. These measurements span two different regions of the nuclear chart in terms of mass and density, and provide crucial test cases for cascade models. Furthermore, the kaonic fluorine measurements explore a regime where the strong interaction induced nuclear capture signature is observable at the measured atomic transitions. To achieve this goal, for the first time, a novel room-temperature Cadmium Zinc Telluride (CZT) detection system, purposely developed by the collaboration for high-resolution kaonic atom X-ray spectroscopy, was deployed in a collider environment.

\noindent Section 2 describes the experimental apparatus. Section 3 presents the data taking, analysis, and Monte Carlo simulations. Section 4 discusses the results, followed by the conclusions, Section 5.

\section{CZT Detectors in the SIDDHARTA-2 Experiment at DA$\Phi$NE}

\noindent The SIDDHARTA-2 experiment \cite{Curceanu:2019uph, Artibani:2024kop, Curceanu:2026zjg} is a dedicated apparatus for precision X-ray spectroscopy of kaonic atoms, installed at the interaction point (IP) of the DA$\Phi$NE electron-positron collider \cite{Milardi:2009zza, Milardi:2018sih, Milardi:2021khj, Milardi:2024efr} at the Laboratori Nazionali di Frascati of the Istituto Nazionale di Fisica Nucleare (INFN-LNF). 

\noindent DA$\Phi$NE operates as a $\phi$-factory, with the centre-of-mass energy tuned to 1.022 GeV, corresponding to the mass of the $\phi$ meson. The $\phi$ meson decays almost at rest into pairs of charged kaons with a branching ratio of 48.9\%. This feature provides a clean and nearly monochromatic source of low-momentum $K^-$ mesons, well suited for the formation of kaonic atoms.

\noindent The SIDDHARTA-2 apparatus is described in detail in \cite{Sirghi:2023wok}. In the present work, only the subsystem relevant to the CZT-based detection of kaonic atom X-rays is discussed, corresponding to the configuration used in this work.

\noindent In the data-taking campaigns, the CZT detection system \cite{abbene_high-rate_2015, abbene_development_2017, abbene_room-temperature_2020} was installed in the plane of the collider rings, in their inner part. The system consists of eight quasi-hemispherical CZT detectors with dimensions of $13 \times 15 \times 5$~mm$^3$, grown using the Traveling Heater Method (THM) by REDLEN Technology (Canada). The detectors are housed in a thin aluminum enclosure equipped with a 0.27 mm-thick aluminum window, which contains both the detector crystals and the front-end electronics. The enclosure was positioned at a distance of 29.3 cm from the collider IP.

\noindent Between the beam pipe and the CZT detection system, a $80 \times 40 \times 2$ mm$^3$ Scionix BC-408 organic scintillator coupled to two photomultiplier tubes via plastic light guides was installed \cite{Skurzok:2020phi}. This detector (LUMI boost), in combination with an identical module placed on the opposite side (LUMI antiboost), was used for luminosity monitoring and for charged kaon selection via time-of-flight techniques as discussed in Section \ref{sec:data analysis}. The scintillator was located at a distance of 10.2 cm from the IP.

\noindent The targets, in which the negative kaons are stopped to form kaonic atoms, consisted of two different material foils, and were placed downstream of the plastic scintillator, facing the detector enclosure. Table \ref{tab:targets} summarizes the thickness and the main characteristics of each target.

\begin{table} [H]

    \centering
    \caption{Target materials together with their densities and thicknesses.}

    \begin{tabular}{cccc}
        Material & Chemical composition & Density (g/cm$^3$) & Thickness (mm) \\
        \hline
        Teflon & C$_2$F$_4$ & 2.2 & 4 \\
        Copper & Cu & 8.96 & 1 \\
    \end{tabular}
    \label{tab:targets}
\end{table}

\begin{figure} [H]
    \centering
    \includegraphics[width=0.75\linewidth]{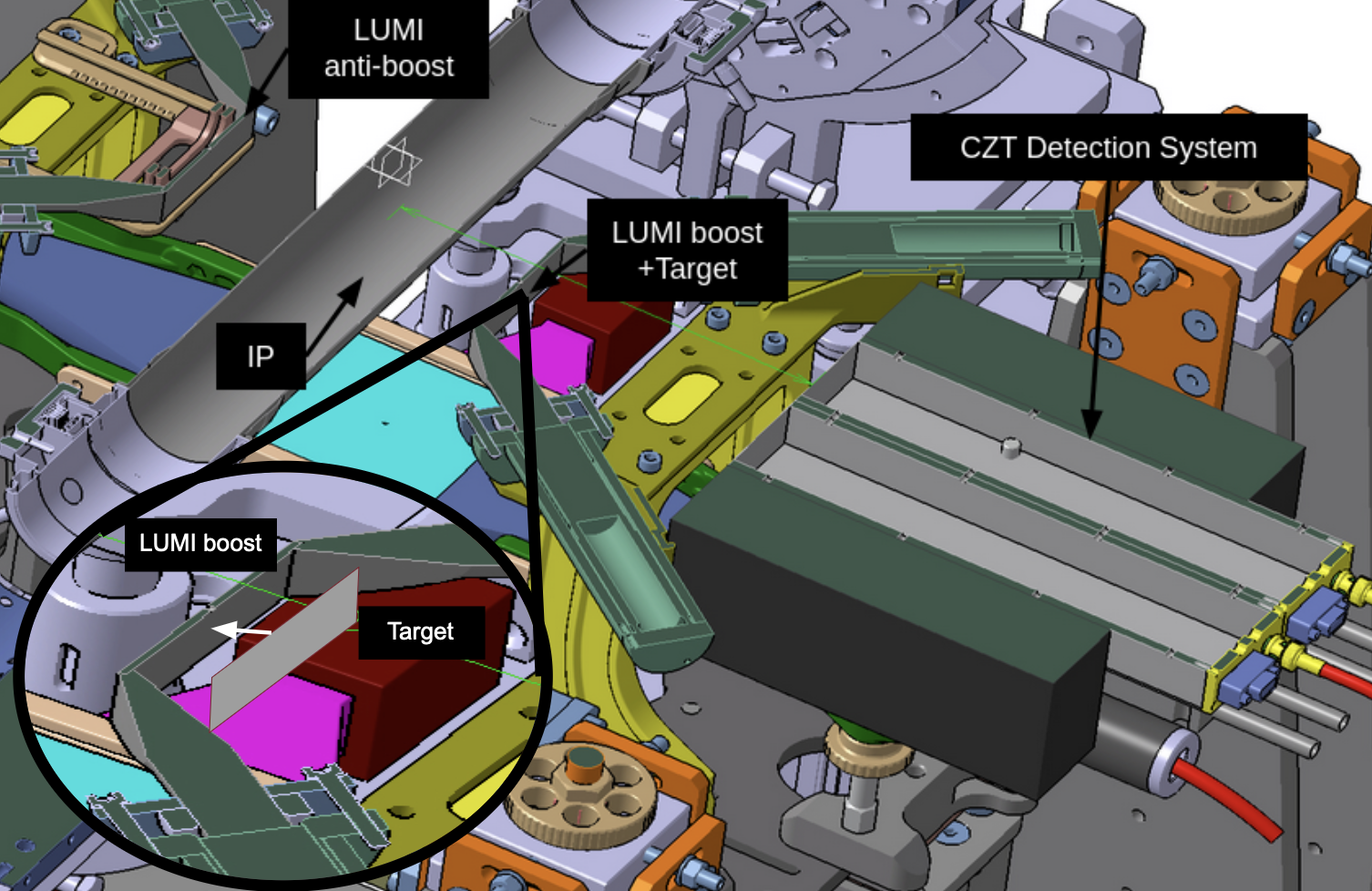}
    \caption{Schematic view of the experimental CZT detection system setup in DA$\Phi$NE used for the data taking presented in this work.}
    \label{fig: CZT setup}
\end{figure}

Figure \ref{fig: CZT setup} represents the experimental setup during all the runs.

\section{Data Taking and Analysis} \label{sec:data analysis}

\noindent Being used for the first time in a collider, the CZT detection system underwent an initial commissioning, testing, and optimization phase in which the stability of the operations \cite{Abbene:2024tpm} and the timing performances \cite{artibani2025Time} were accurately tested. As a result the number of CZT detectors actively operating and used for physics analysis varied across the data-taking.

\noindent Table \ref{tab:runs} summarizes the 2024 data-taking periods considered in the present analysis, together with the corresponding delivered integrated luminosity and the number of CZT detectors included in the data analysis for each target.

\begin{table} [H]

    \centering
    \caption{Summary of the runs used in the data analysis for each target.}
    
    \begin{tabular}{ccc}
        Target  &  Delivered Integrated   & Active Detectors    \\
              &  Luminosity (pb$^{-1}$) & (for Data Analysis) \\
        \hline
         \multirow{2}{*}{Cu}
         & 35 & 8 \\
         & 66 & 7 \\
         \hline
         \multirow{2}{*}{Teflon}
         & 33 & 6 \\
         & 41 & 6 \\
         \hline
    \end{tabular}
    \label{tab:runs}
\end{table}

\subsection{Data Selection}

\begin{figure} [H]
    \centering
    \includegraphics[width=1\linewidth]{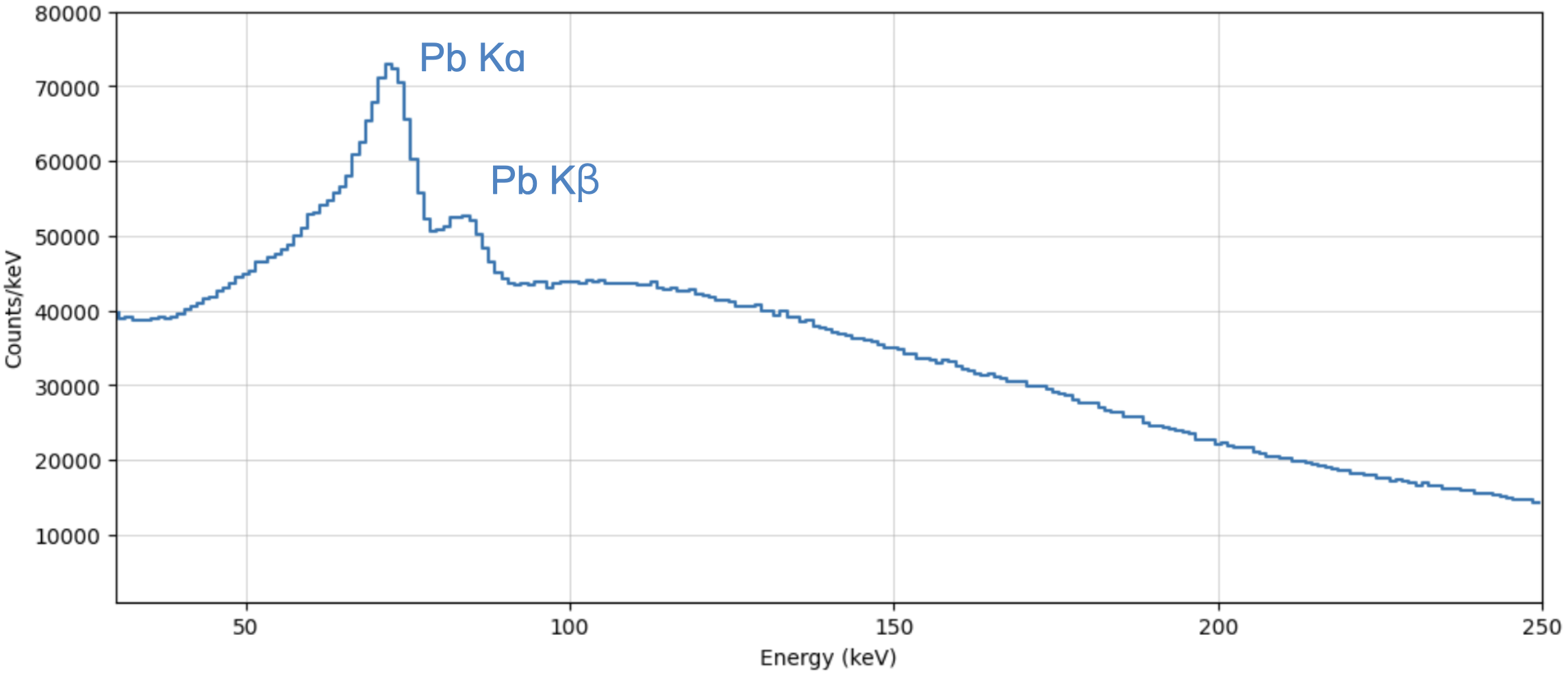}
    \caption{Spectrum of triggered events collected during the second run with copper target.}
    \label{fig: raw spectrum}
\end{figure}

\noindent The events are selected by requiring a trigger coincidence between the two luminometers and the detector within a 10 $\mu$s time window. For each triggered event the full CZT detectors waveform is recorded. A typical raw energy spectrum corresponding to one of the data-taking periods is shown in Figure \ref{fig: raw spectrum}.

\noindent Lead fluorescence lines (75 keV and 85 keV) are observed over the background. To extract X-rays originating from kaonic
atom transitions, a dedicated data-selection procedure, based on timing information from both the luminometers and the CZT detectors, was applied as described in \cite{artibani2025Time}.

\noindent Since the primary goal of this work is the measurement of absolute yields, conservative and uniform selection criteria were applied to all datasets, in order to maximise efficiency over purity. The kaon time selection window was chosen to be 1 ns wide, while a time window of 170 ns width was applied to the CZT-luminometer time difference distribution for all targets and data-taking periods (see \cite{artibani2025Time} for technical details). 


\subsection{Fit Functions}

\noindent The final energy spectra, calibrated following the procedure described in \cite{Abbene:2024tpm}, were summed target-by-target and subsequently analysed to extract the energy and the number of events for each kaonic atoms X-ray transitions.

\noindent The background contribution was modelled using the function

\begin{equation}
    f_{{bkg}}(x) = a + b\,x + c\,\exp(d\,x) + \text{erfc}\!\left(\frac{e - x}{f}\right),
\end{equation}

\noindent where the linear part ($a$ and $b$ parameters) describes the electronics baseline, the exponential part ($c$ and $d$ parameters) describes background generated by photons and electrons undergoing Compton scattering, and the complementary error function part ($e$ and $f$ parameters) accounts for the low-energy cutoff resulting from the presence of shielding and electronic components. This background was previously validated in \cite{Abbene:2024tpm} using beam-on data with a known radioactive source.

\noindent The X-ray peaks corresponding to kaonic atom transitions were described by Gaussian functions,

\begin{equation} \label{eq_peak}
    f_{{peak}}(x) = N \exp \left(-\frac{(x-\mu)^2}{2\sigma^2}\right),
\end{equation}

\noindent where $N$ is the peak area, $\mu$ the centroid energy, and $\sigma$ the detector resolution.

\noindent For the 75 keV peak arising from fluorescence of the lead shielding after activation, an additional low energy tail component accounting for incomplete charge collection in the CZT detectors was included. The tail was modeled as

\begin{equation}
f_{{tail}}(x) =
\epsilon\, N \exp \left(\frac{x-\mu}{\beta \sigma}\right)
{erfc} \left(
\frac{x-\mu}{\sqrt{2}\sigma} + \frac{1}{\sqrt{2}\beta}
\right),
\end{equation}

\noindent where $\epsilon$ represents the relative intensity of the tail and $\beta$ its characteristic width.

\noindent For all other fitted peaks, the tail contribution was found to be statistically insignificant and was therefore omitted.

\noindent To account for the intrinsic resolution of the detector, the $\sigma$ parameters of the Gaussian, were described as a function of the energy, following the formula
\begin{equation}
    \sigma(x) = \sqrt{a_{\sigma} \cdot x + b_{\sigma}}
\end{equation}

\noindent with $a_{\sigma}$ and $b_{\sigma}$ free parameters common to all the peaks.

\subsection{Fit Method}

\noindent The two spectra corresponding to copper and teflon target were simultaneously fitted, by means of $\chi^2$ minimisation, with the following common parameters:

\begin{itemize}
    \item The Gaussian $\sigma$ parameters $a_{\sigma}$ and $b_{\sigma}$;
    
    \item The mean of the lead K$\alpha$ and K$\beta$ fluorescence peaks;

    \item The relative amplitude of the lead K$\beta$ with respect to the K$\alpha$ fluorescence peak;

    \item The tail parameters $\epsilon$ and $\beta$ for the lead K$\alpha$ fluorescence peak;
\end{itemize}

\noindent while the background and kaonic atom transition parameters were kept independent for each dataset. The fit was performed in this way  to have a better description of the background due to the lead fluorescence (Pb K$\alpha$ and Pb K$\beta$ fluorescence lines), and of the parameters of the energy resolution.


\subsection{Results}

\begin{figure} []
    \centering
    \includegraphics[width=1\linewidth]{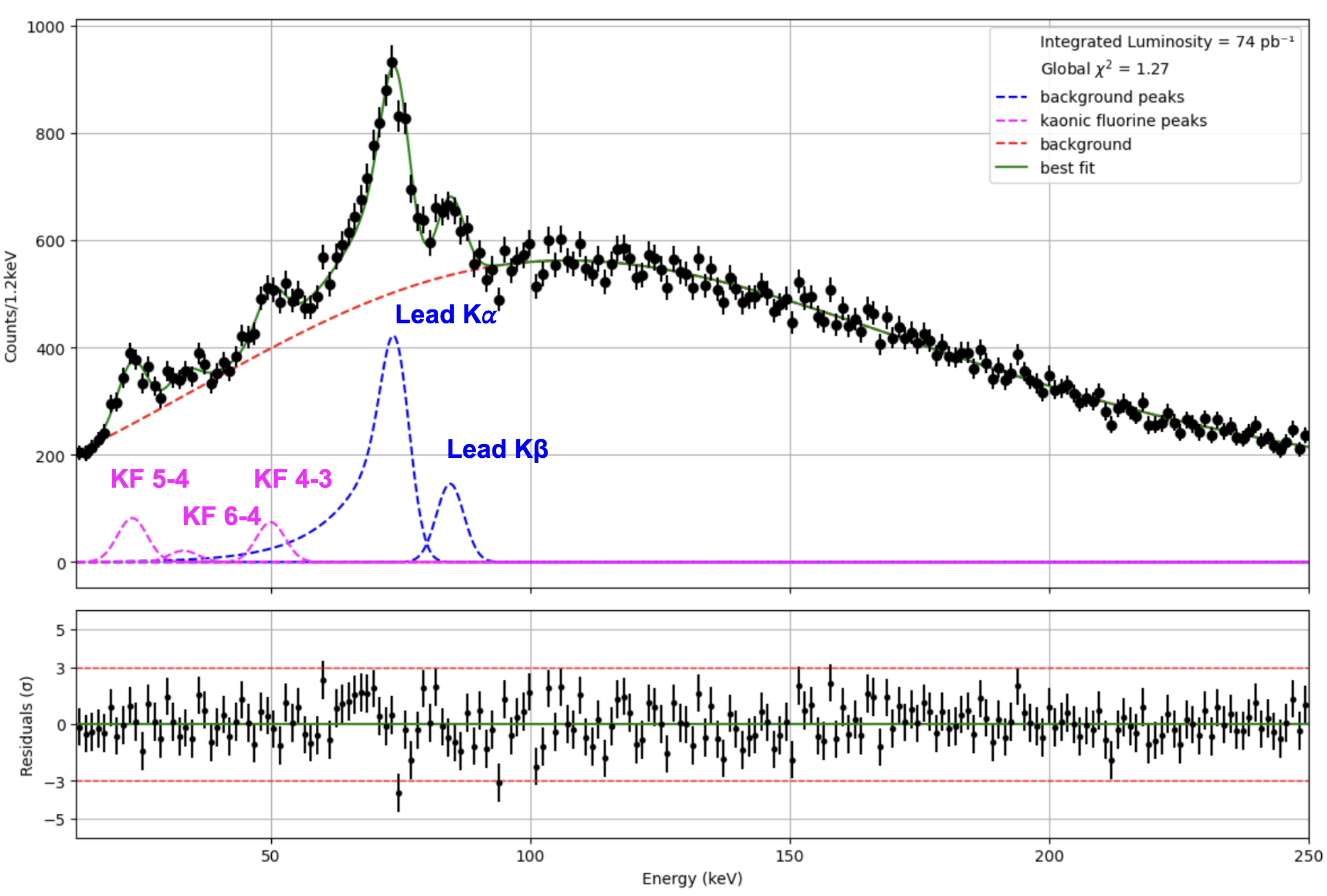}
    \caption{Upper: Kaonic fluorine energy spectrum and fit. The experimental counts are shown in black, with error bars representing the statistical uncertainty on the bin counts. The contributions from lead fluorescence transitions are shown in blue, kaonic atom transitions in magenta, the background in dashed red, and the total fit in green. Lower: Relative residuals of the fit.}
    \label{fig: fluorine fit}
\end{figure}

\begin{figure} []
    \centering
    \includegraphics[width=1\linewidth]{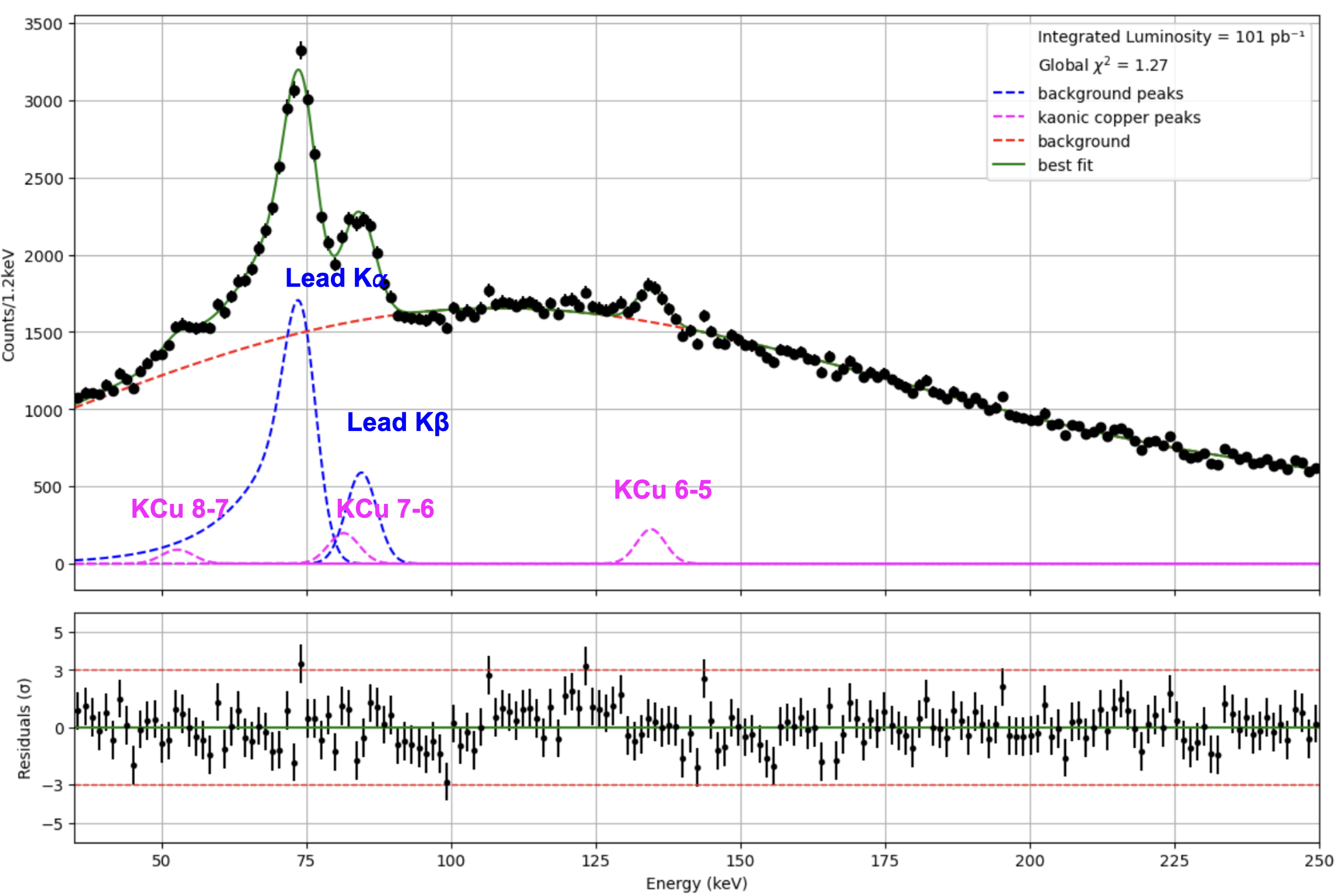}
    \caption{Upper: Kaonic copper energy spectrum and fit. The experimental counts are shown in black, with error bars representing the statistical uncertainty on the bin counts. The contributions from lead fluorescence transitions are shown in blue, kaonic atom transitions in magenta, the background in dashed red, and the total fit in green. Lower: Relative residuals of the fit.}
    \label{fig: copper fit}
\end{figure}

\noindent The results of the fit are presented in Figures \ref{fig: fluorine fit} and \ref{fig: copper fit}. The reduced chi-squared of the simultaneous fit resulted to be $\chi^2_{red} = 1.27$.

\noindent The measured energies and the corresponding numbers of events for the observed kaonic atom transitions, as obtained from the fits, are summarized in Table \ref{tab:energies_events}.

\begin{table} [H]
    \centering
    \caption{Energy and number of events extracted from the fits, with the corresponding theoretical energies calculated using the MCDFGME code \cite{santos_x-ray_2005}}

    \begin{tabular}{cccc}
    \multirow{2}{*}{Transition} & Measured Energy                & \multirow{2}{*}{Theoretical Energy (keV)} & Number of Events \\
                                & $\pm$ stat. $\pm$ sys. (keV)   &                                           & $\pm$ stat. $\pm$ sys. \\
    \hline
    \multicolumn{4}{c}{Fluorine} \\
    \hline
    KF 5 $\rightarrow$ 4 & 23.34 $\pm$ 0.34 $\pm$ 0.09 & 23.33 & 488 $\pm$ 72 $\pm$ 30 \\
    KF 4 $\rightarrow$ 3 & 50.14 $\pm$ 0.54 $\pm$ 0.15 & 50.59 & 427 $\pm$ 74 $\pm$ 42 \\
    \hline
    \multicolumn{4}{c}{Copper} \\
    \hline
    KCu 8 $\rightarrow$ 7 & 52.89 $\pm$ 0.50 $\pm$ 0.30 & 52.63 & 578 $\pm$ 84 $\pm$ 38 \\
    KCu 7 $\rightarrow$ 6 & 80.89 $\pm$ 0.98 $\pm$ 0.60 & 81.17 & 1078 $\pm$ 297 $\pm$ 190 \\
    KCu 6 $\rightarrow$ 5 & 134.48 $\pm$ 0.35 $\pm$ 0.31 & 134.81 & 1206 $\pm$ 132 $\pm$ 42 \\
    \end{tabular}
    \label{tab:energies_events}
\end{table}

\noindent The systematic uncertainties were evaluated by repeating the fitting procedure under different analysis conditions, following the approach already adopted in \cite{artibani2025Time}.

\subsection{Monte Carlo Simulation}

\noindent In order to account for the geometrical acceptance and the detector response, and to determine the number of kaons stopped in the target, essential ingredients for the extraction of absolute X-ray yields per stopped kaon, a Monte Carlo simulation based on the GEANT4 toolkit was developed. The simulation reproduces the experimental setup, including the exact geometry of the CZT crystals and their spacing, the aluminium entrance window, the luminometer system, and the DA$\Phi$NE beam pipe.

\noindent The simulation starts with the production of a $\phi$ meson at the interaction point, with a momentum distribution boosted toward the detector side, consistently with the DA$\Phi$NE machine parameters. The subsequent $\phi$ decays, including the decay into charged kaon pairs, are simulated, and the transport of the kaons through the experimental apparatus is tracked until their decay or stopping in the target or surrounding materials.

\noindent Whenever a negatively charged kaon stops in the target material, and forms a kaonic atom, the X-ray transitions of interest within the energy range of the CZT detectors are isotropically generated with a radiative yield set to 100\% (i.e. used as a normalization reference for efficiency determination). This choice provides a normalization reference, allowing the experimental yields to be extracted by comparison with the simulated X-ray intensities.

\noindent Being teflon a compound element (C$_2$F$_4$), kaons stopped in the material can bind to either fluorine or carbon nuclei, and the exact capture process remains uncertain. In Reference \cite{Wiegand:1969cu}, kaons were assumed to stop on the elements of a compound in proportion to the atomic charge fraction, which for fluorine gives a factor $f_Z = Z_F\cdot N_F/(Z_F\cdot N_F+Z_C\cdot N_C) = 0.75$, with $Z_F$ and $Z_C$ atomic number of fluorine and carbon, and $N_F$ and $N_C$ atomic abundance in teflon respectively for fluorine and carbon. This approach was originally suggested by Fermi and Teller in their study of exotic atoms, but significant deviations from this behaviour were found in muonic and pionic atoms \cite{Wiegand:1973ac}. The SIDDHARTA collaboration, in measurements on kapton (C$_{22}$H$_{10}$N$_2$O$_5$) \cite{SIDDHARTA:2013ftw}, used instead a factor proportional to the number of nuclei in the molecule, which for fluorine in teflon gives $f_N = N_F/(N_F+N_C) = 0.66$. Given these theoretical uncertainties, it was decided to generate an X-ray every time a kaon stops in the compound, and to correct the result a posteriori using both factors.

\subsection{Yields Measurement}

\noindent Following the approach adopted in previous kaonic atom experiments \cite{Sgaramella:2023jsc}, the absolute X-ray yield $Y$ per stopped kaon for a given transition is defined as the ratio between the experimental and simulated number of X-rays from kaonic atoms, normalized to the kaon pairs detected by the luminometers.

\begin{equation}
    Y_{rad} = \frac{N^{{EXP}}_{{X\text{-}ray}} / N^{{EXP}}_{{KLUMI}}}
             {N^{{MC}}_{{X\text{-}ray}} / N^{{MC}}_{{KLUMI}}} ,
\end{equation}

\noindent where $N^{{EXP}}_{{X\text{-}ray}}$ and $N^{{EXP}}_{{KLUMI}}$ denote, respectively, the number of X-ray events observed for the selected kaonic atom transition and the number of kaons measured by the luminometer in the experimental data. The quantities $N^{{MC}}_{{X\text{-}ray}}$ and $N^{{MC}}_{{KLUMI}}$ are the corresponding values obtained from the Monte Carlo simulation.

\noindent The systematic errors from the Monte Carlo simulation were evaluated by moving the target from its nominal position (just downstream of the LUMI) by 0.5 mm, varying the target thickness by $\pm 0.1$ mm, and including the statistical uncertainty on the simulated counts.

\noindent In Table \ref{tab: yield copper} the measured yields for kaonic copper are presented.

\begin{table} [H]
    \centering
     \caption{Kaonic copper absolute yields per kaon stopped, with the corresponding statistical and systematic error. \\
    * The KCu 6 $\rightarrow$ 5 transition in the measurement is contaminated by the KCu 8 $\rightarrow$ 6 transition. This contribution was considered in the systematics (see text).}
    
    \resizebox{140pt}{!}{
    \begin{tabular}{lc}
    
    Transition &  Yield (\%) \\
    \hline \\[-5pt]
        KCu 8 $\rightarrow$ 7 & 17.6 $\pm$ 2.6 $^{+ 1.5}_{- 1.5}$ \\ [5pt]
        KCu 7 $\rightarrow$ 6 & 18.5 $\pm$ 5.1 $^{+ 3.0}_{- 3.0}$ \\ [5pt]
        KCu 6 $\rightarrow$ 5 *  & 21.4 $\pm$ 2.3 $^{+ 1.3}_{- 3.0}$ \\ [5pt]
    \hline
    \end{tabular}
    }
    \label{tab: yield copper}
\end{table}

\noindent In kaonic copper, as in several other kaonic atoms \cite{Wiegand:1973ac}, the energies of the $6 \rightarrow 5$ transitions overlap with those of the $8 \rightarrow 6$ transitions, making them experimentally indistinguishable for a yield measurement. Nevertheless, the $\Delta n = 2$ transitions are disfavoured by selection rules, and their contribution to the total yield is approximately 10\% of that of the principal $\Delta n = 1$ transition \cite{Wiegand:1973ac}. In the present case, the latter is the $8 \rightarrow 7$ transition, measured to be 17.6\%. Consequently, the contribution of the $8 \rightarrow 6$ transition to the overall yield of the $6 \rightarrow 5$ transition is expected to be of the order of 2-3\%, considered in the systematic error.

\noindent In Table \ref{tab: yield res} the measured yields for kaonic fluorine are presented.

\begin{table} [H]
    \centering
     \caption{Kaonic fluorine absolute yields, with the corresponding statistical error and systematic error}
    
    \resizebox{\textwidth}{!}{
    \begin{tabular}{lccc}
    
    \multirow{2}{*}{Transition} & Yield in Teflon (\%) & Yield in F with $f_Z = 0.75$ (\%) & Yield in F with $f_N = 0.66 $  (\%) \\
    & $\pm$ stat. $\pm$ sys.  & $\pm$ stat. $\pm$ sys. & $\pm$ stat. $\pm$ sys. \\
    \hline
    KF 5 $\rightarrow$ 4 & 9.8 $\pm$ 1.5 $\pm$ 0.6 & 13.1 $\pm$ 2.0 $\pm$ 0.9 & 14.8 $\pm$ 2.3 $\pm$ 1.1 \\
    KF 4 $\rightarrow$ 3 & 7.5 $\pm$ 1.3 $\pm$ 0.8 & 10.0 $\pm$ 1.7 $\pm$ 0.9 & 11.4 $\pm$ 1.9 $\pm$ 1.1 \\
    \hline
    \end{tabular}
    }
    \label{tab: yield res}
\end{table}

\noindent The measured yield per kaon stopped in copper for the 8 $\rightarrow$ 7 and 6 $\rightarrow$ 5 in Table \ref{tab: yield copper} represent the new most precise measurement, and together with the 7 $\rightarrow$ 6 show a good compatibility with the old measurement done in \cite{Wiegand:1973ac}, while the yield for the transitions in Table \ref{tab: yield res} represent the first measurement ever.

\section{Discussion}

\noindent The intensity of the spectral lines, and consequently the X-ray yields per stopped kaon, produced by kaonic atoms as a function of the principal quantum number result from a complex interplay between radiative transitions, Auger de-excitation from the electronic shells, and nuclear capture induced by the strong interaction.

\noindent From cascade considerations \cite{Koike:2007uye}, the radiative transition rate increases as the principal quantum number $n_i$ decreases, while the Auger rate shows the opposite trend, becoming more important at higher $n_i$. The nuclear capture probability associated with the strong interaction, on the other hand, remains negligible at large atomic radii and starts to increase only when the kaon wave function significantly overlaps with the nuclear density. Within this framework, the experimental X-ray yields encode the cumulative effect of these competing processes.

\noindent For kaonic copper, this trend is confirmed by the observations in Table \ref{tab: yield res}: the yields increase smoothly as $n_i$ decreasing, consistent with a radiative-dominated cascade in which Auger processes lose importance, and nuclear capture contribution is negligible.

\noindent A different behaviour is observed for kaonic fluorine. The yield of the $4\rightarrow3$ transition is suppressed with respect to the higher-$n$ transitions, suggesting the onset of strong interaction effects at the $n=4$ level.

\begin{figure}[H]
    \centering
    \includegraphics[width=1\linewidth]{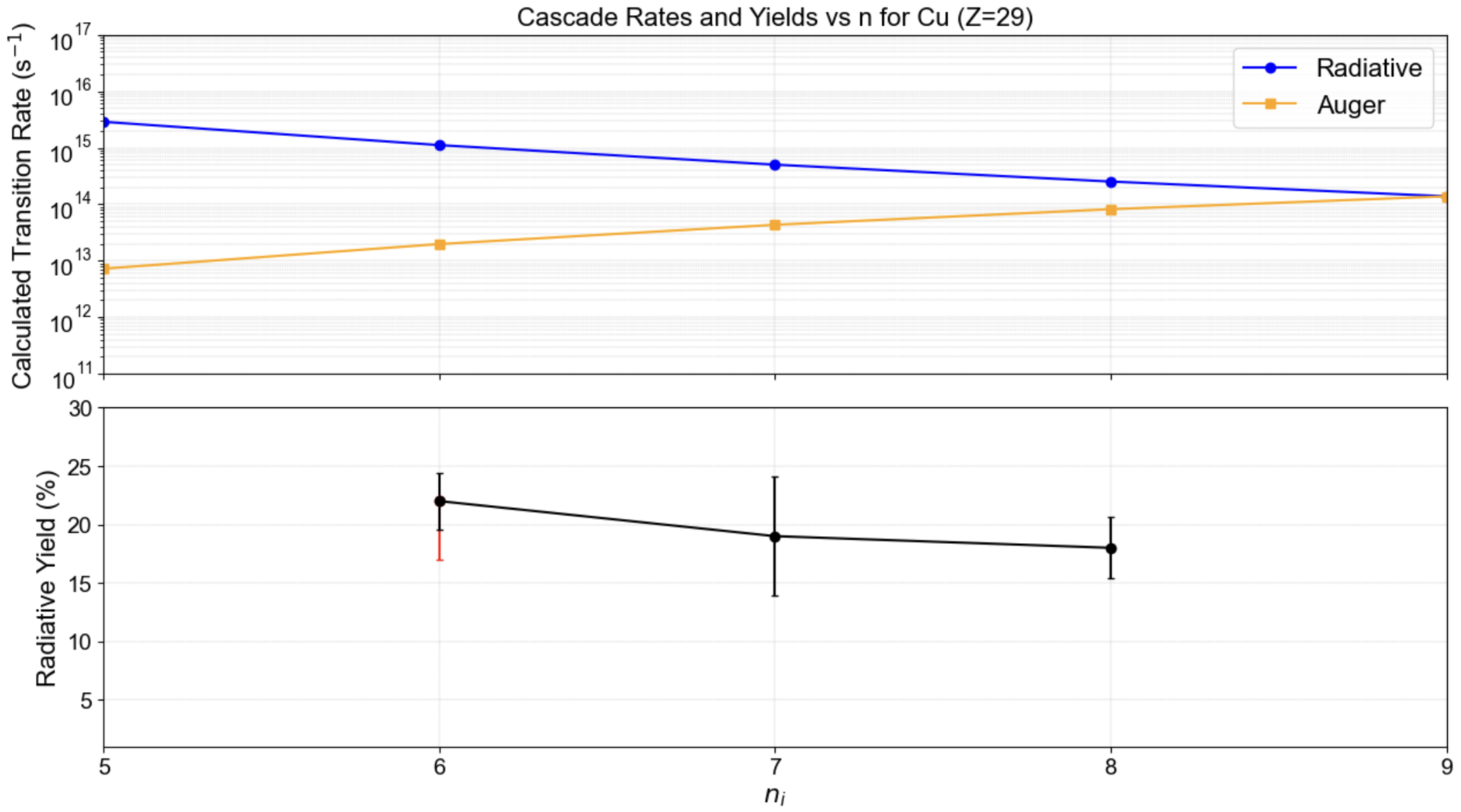}
    \caption{Upper panel: radiative (blue) and Auger K-shell (orange) transition rates for circular states ($l=n-1$) with $\Delta n = 1$. A conservative systematic uncertainty of $-5\%$ (red) was assigned to the 6 $\rightarrow$ 5 transition yield due to the uncertainty in the 8 $\rightarrow$ 6 transition's yield (see text).}
    \label{fig: yield rates KCu}
\end{figure}

\begin{figure}[H]
    \centering
    \includegraphics[width=1\linewidth]{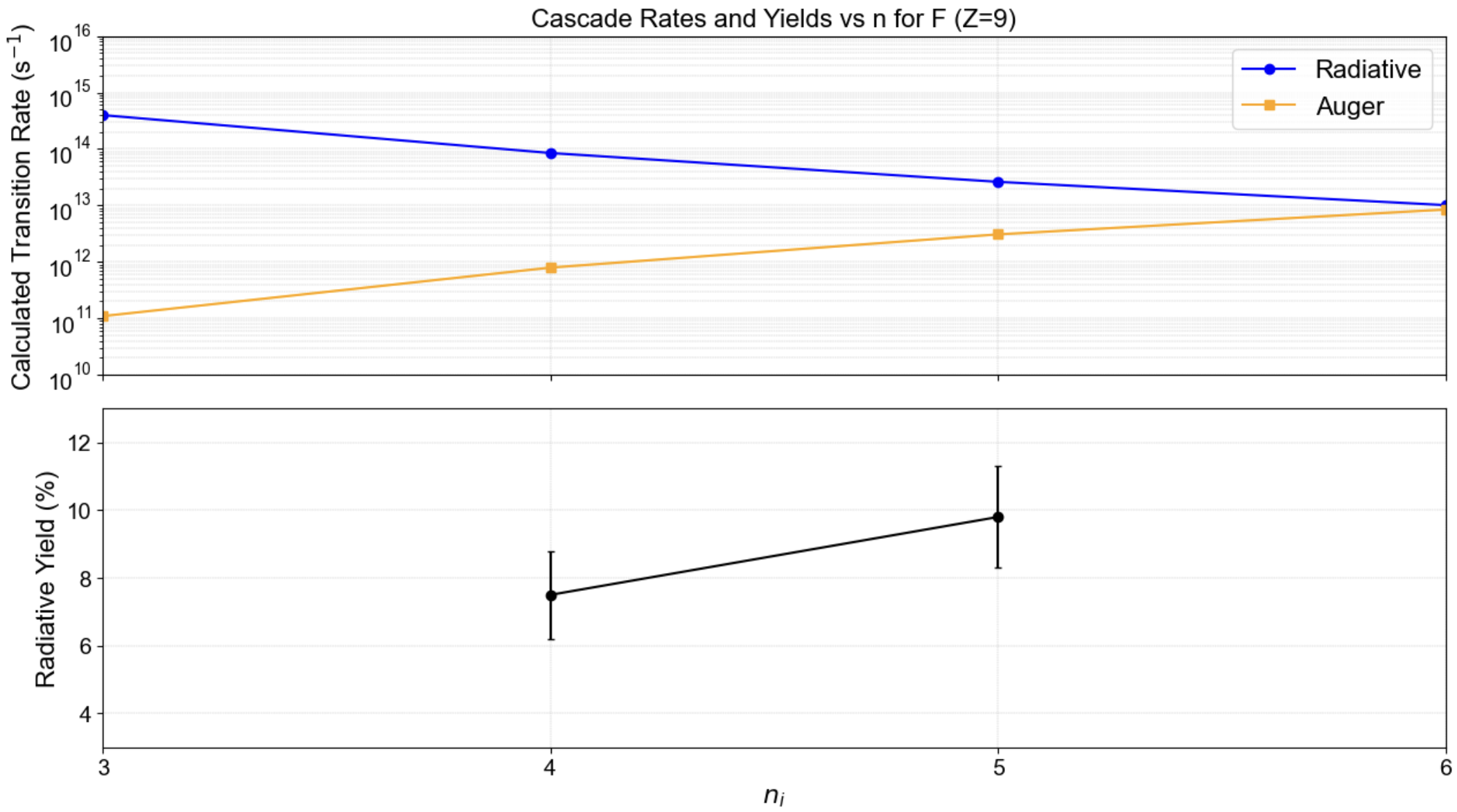}
    \caption{Upper panel: radiative (blue) and Auger K-shell (orange) transition rates for circular states ($l=n-1$) with $\Delta n = 1$ starting from level $n_i$ in kaonic fluorine. Lower panel: measured X-ray yield per stopped kaon.}
    \label{fig: yield rates KF}
\end{figure}

\noindent To support this interpretation, Figures \ref{fig: yield rates KCu} and \ref{fig: yield rates KF} show, for kaonic copper and kaonic fluorine, respectively, the calculated radiative and Auger K-shell rates for circular transitions ($\Delta n = 1$, $l = n-1$) starting from level $n_i$, and the measured X-ray yields.


\noindent Following \cite{BurbidgePhysRev.89.189}, as done in \cite{SIDDHARTA-2:2025mwc}, the radiative rates $\Gamma_{rad}$ were derived using Equation \ref{Eq: gamma rad}, assuming electric dipole transitions, hydrogenic wave functions with reduced mass $\mu$:

\begin{equation} \label{Eq: gamma rad}
    \Gamma^{rad}_{(n,l) \rightarrow (n', l\pm1)} = \frac{4\mu Z^4 \alpha^3}{3} \cdot \left| \mathcal{R}^{(n,l)}_{(n', l\pm1)} \right|^2 \cdot \left( \Delta E_{(n,n')} \right) ^3,
\end{equation}

\noindent where $Z$ is the atomic number, $\mu$ the reduced mass of the exotic atom system, $\alpha$ the fine structure constant, and $\mathcal{R}^2$ is the dipole radial integral evaluated analytically.

\noindent The Auger K-shell rate was estimated assuming a single K-shell electron with effective nuclear charge $Z_e = Z - 1$ with the following formula:

\begin{equation} \label{Eq Aug1s}
    \Gamma^{Aug,1s}_{(n,l) \rightarrow (n', l\pm1)} = \frac{16}{3} \cdot \frac{\pi Z_e^2}{\mu^2 Z^2} \cdot
    \frac{l}{2l+1}
    \cdot \left| \mathcal{R}^{(n,l)}_{(n', l\pm1)} \right|^2 \cdot \frac{y_K^2}{1 + y_K^2}
    \cdot \frac{\exp\left(y_K(4\tan^{-1} y_K - \pi)\right)}{\sinh(\pi y_K)},
\end{equation}

\noindent where $\mathcal{R}^2$ is the dipole radial integral, and $y_K$ is the Sommerfeld parameter:

\begin{equation}
    y_K = \frac{Z_e}{\sqrt{2 E_K}}
\end{equation}

\noindent with $E_K$ is the electron energy in the K shell.

\noindent Looking at Figure~\ref{fig: yield rates KF}, the observed suppression of the $4 \rightarrow 3$ yield in kaonic fluorine suggests that nuclear capture already contributes at $n=4$. The strong interaction width $\Gamma_{\text{strong}}$ for a level $n$ can be estimated from the radiative yield attenuation, assuming negligible Auger rates~\cite{Ericson:1970ij}:

\begin{equation}
    \Psi_n = \frac{\Gamma_{\text{rad}}^{(n)}}{\Gamma_{\text{rad}}^{(n)} + \Gamma_{\text{strong}}^{(n)}} = 
    \frac{Y_{n \rightarrow n-1;rad}}{\sum_{N>n} (Y_{N \rightarrow n;rad} + (Y_{N \rightarrow n;Aug})},
\end{equation}

\noindent where the sum in the denominator runs over all transitions feeding level $n$ from higher-lying states. A determination of $\Gamma_{\text{strong}}$ would therefore require a full cascade calculation to evaluate this feeding contribution.

\noindent In the absence of a dedicated cascade model for kaonic fluorine, we instead use the measured X-ray yields per stopped kaon to set a constraint on $\Gamma_{\text{strong}}$ for $n = 4$. From the experimental data we obtain the ratio
\begin{equation}
R = \frac{Y_{4 \to 3 ; rad}}{Y_{5 \to 4;rad}} = 0.77 \pm 0.17.
\end{equation}

\noindent The relative yield of the $4 \to 3$ transition includes not only direct population from the $5 \to 4$ transition, but also additional feeding from higher levels (e.g. $6 \to 4$, $7 \to 4$, \dots), which are not accounted for in the denominator. As a consequence, the ratio $R$ overestimates the true radiative yield of the $n = 4$ level,

\begin{equation}
\Psi_4 = \frac{\Gamma^{(4)}_{\text{rad}}}{\Gamma^{(4)}_{\text{rad}} + \Gamma^{(4)}_{\text{strong}}},
\end{equation}
implying $R \geq \Psi_4$. This leads to the inequality
\begin{equation}
\Gamma^{(4)}_{\text{strong}} \geq \Gamma^{(4)}_{\text{rad}} \left( \frac{1}{R} - 1 \right).
\end{equation}

\noindent To obtain a conservative lower limit, we consider the maximum value of $R$ within one standard deviation, $R_{\text{max}} = 0.94$, yielding

\begin{equation}
    \Gamma_{\text{strong}}^{(4)} > 0.056 \cdot \left( \frac{1}{0.94} - 1 \right) = 0.0036~\text{eV} \quad (90\%\ \text{C.L.}),
\end{equation}

\noindent with $\Gamma^{(4)}_{\text{rad}} = 0.056$ eV, calculated using Equation \ref{Eq: gamma rad}.

\noindent This lower limit is compatible with the theoretical estimate of $\approx 0.006$~eV, obtained by theoretical calculation based on \cite{Obertova:2022des, Obertova:2025rso}.

\noindent The present work also provides a quantitative constraint that will allow future cascade calculations to be tested against data through new precise measurement of kaonic copper and fluorine transition not affected by strong interaction.

\section{Conclusions}

\noindent The present work reports new measurements of absolute X-ray yields for various transitions in kaonic copper and fluorine. For kaonic copper, the two most precise measurements to date are obtained for the $8\rightarrow7$ and $6\rightarrow5$ transitions, while for kaonic fluorine the absolute yields of the $5\rightarrow4$ and $4\rightarrow3$ transitions are reported for the first time. These results provide key observables that are sensitive to the atomic cascade dynamics and to the nuclear absorption.

\noindent In particular, a reduction of the yield between consecutive transitions is observed in kaonic fluorine suggesting the onset of nuclear capture at the $n=4$ level, that led to the evaluation of a lower limit for the strong interaction in this level.

\noindent From this perspective, the present measurements complement traditional spectroscopic observables such as level shifts and widths induced by the strong interaction. With a reliable cascade model, a quantitative determination of the nuclear capture rate at these levels becomes possible, probing a regime inaccessible to shift and width measurements alone.

\noindent The measured yields also provide valuable inputs for cascade model development, in particular for constraining the role of electron refilling and the density of states during the de-excitation process, with implications across a broad range of exotic atom studies.

\noindent Finally, the present results establish CZT-based detection systems as a mature technology for high-precision kaonic atom spectroscopy, providing the experimental precision needed for yield extraction at the levels required by cascade models. This work opens a new experimental window onto the interplay of atomic cascade dynamics and nuclear absorption in kaonic atoms, establishing absolute yield measurements as a crucial experimental probe of the kaon-nucleon interaction at low energies.

\section*{Acknowledgements}
{
\noindent The authors gratefully acknowledge Riccardo Gargana for access to computational resources. We thank Jaroslava Obertova for estimating the strong interaction shifts and widths. We thank C. Capoccia from INFN-LNF for his fundamental contribution in designing and building the SIDDHARTA-2 setup. We thank as well the INFN, INFN-LNF and the DA$\Phi$NE staff in particular to Dr. Catia Milardi for the excellent working conditions and permanent support.
}

{\noindent
Part of this work was supported by the Austrian Science Fund (FWF): [P24756-N20 and P33037-N]; the Croatian Science Foundation under the project HRZZ-IP-2022-10-3878; the EU STRONG-2020 project (Grant Agreement No. 824093); the EU Horizon 2020 project under the MSCA (Grant Agreement 754496); the Japan Society for the Promotion of Science JSPS KAKENHI Grant No. JP18H05402; the SciMat and qLife Priority Research Areas budget under the program Excellence Initiative - Research University at the Jagiellonian University, and the Polish National Agency for Academic Exchange (Grant No. PPN/BIT/2021/1/00037); the EU Horizon 2020 research and innovation programme under project OPSVIO (Grant Agreement No. 101038099). This work was also supported by the Italian Ministry for University and Research (MUR), under PRIN 2022 PNRR project CUP: B53D23024100001. This article/publication is based upon work from COST ActionCA24131-ENRICH, supported by COST (European Cooperation in Science and Technology, http://www.cost.eu/).
}


\bibliographystyle{elsarticle-num} 
\bibliography{ref.bib}

\end{document}